\documentclass[a4paper,amsfonts, amssymb, amsmath,preprint, showkeys, nofootinbib, twoside]{revtex4-1}
\usepackage[english]{babel}
\usepackage{ulem}
\usepackage[utf8]{inputenc}
\usepackage[colorinlistoftodos, color=green!40, prependcaption]{todonotes}
\usepackage{amsthm}
\usepackage{mathtools}
\usepackage{physics}
\usepackage{xcolor}
\usepackage{graphicx}
\usepackage[left=23mm,right=13mm,top=35mm,columnsep=15pt]{geometry}
\usepackage{adjustbox}
\usepackage{placeins}
\usepackage[T1]{fontenc}
\usepackage{lipsum}
\usepackage{csquotes}
\usepackage{enumerate}
\usepackage{epstopdf}
\usepackage{makeidx}
\usepackage{caption}
\usepackage{subcaption}
\usepackage{fancyhdr}
\usepackage{marvosym}
\usepackage{wrapfig}
\usepackage{indentfirst} 
\def\bef{\begin{framed}}
\def\eef{\end{framed}}
\def\be{\begin{equation}}
\def\ee{\end{equation}}
\def\ber{\begin{eqnarray}}
\def\eer{\end{eqnarray}}

\def\sigmabold{\mbox{\boldmath $\sigma$}}
\def\thetabold{\mbox{\boldmath $\theta$}}

\def\ev{{\bf e}}

\def\zv{{\bf \hat z}}

\def\jv{{\bf j}}
\def\Jv{{\bf J}}
\def\kv{{\bf k}}

\def\Av{{\bf A}}

\def\Ev{{\bf E}}

\def\Rv{{\bf R}}

\def\Tv{{\bf T}}

\def\nn{\nonumber}

\def\Bc{{\mathcal{B}}}
\def\Hc{{\mathcal{H}}}

\def\Mc{{\mathcal{M}}}

\usepackage[pdftex, pdftitle={Article}, pdfauthor={Author}]{hyperref} 
\begin{document}
\title{Frequency-dependent Faraday and Kerr rotation in anisotropic nonsymmorphic Dirac semimetals}

\author{Amarnath Chakraborty}
\email{achakraborty@mail.missouri.edu}
\affiliation{Department of Physics and Astronomy, University of Missouri, Columbia, Missouri, USA}
\author{Guang Bian}
\affiliation{Department of Physics and Astronomy, University of Missouri, Columbia, Missouri, USA}
\author{Giovanni Vignale}
\email{vignaleg@missouri.edu}
\affiliation{Department of Physics and Astronomy, University of Missouri, Columbia, Missouri, USA}
\date{\today} 
\begin{abstract}
We calculate the frequency-dependent longitudinal and Hall conductivities and the Faraday and Kerr rotation angles for a single sheet of anisotropic Dirac semimetal protected by nonsymmorphic symmetry in the presence of a Zeeman term coupling to the out-of-plane component of the spin.  While the Zeeman term causes a rotation of the plane of polarization of the light, the anisotropy causes the appearance of an elliptically polarized component in an initially linearly polarized beam. The two effects can be combined in a single complex Faraday rotation angle. At the zero-frequency limit, we find a finite value of the Faraday rotation angle, which is given by $2\alpha_F$, where $\alpha_F$ is the effective fine structure constant associated with the velocity of the linearly dispersing Dirac fermions. We also find a logarithmic enhancement of the Faraday (and Kerr) rotation angles as the frequency of the light approaches the absorption edge associated with the Zeeman-induced gap. While the enhancement is reduced by impurity scattering, it remains significant for an attainable level of material purity. These results indicate that two-dimensional Dirac materials protected by nonsymmorphic symmetry are responsive to Zeeman couplings and can be used as platforms for magneto-optic applications, such as the realization of polarization-rotating devices.

\end{abstract}


\maketitle

\section{Introduction}

Two-dimensional (2D) electronic systems have garnered tremendous attention over the past decade due to their exceptional optoelectronic properties and gate-tunable response \cite{Bonaccorso2010, Mak2016}. Starting with graphene \cite{Novoselov2005}, followed by Dirac and Weyl semimetals, these materials~\cite{3D, Yan, Nagaosa2020} have been shown to possess several unique electronic and optical properties, which are traceable to their linear energy dispersion and non-trivial Berry phase. 

However, the Dirac points in many existing 2D materials, including graphene, are vulnerable to spin-orbit coupling (SOC). Motivated by finding alternative 2D materials beyond graphene, various atomically thin materials,  including silicene, MoS$_2$, and phosphorene, have been added to the list, each with its peculiar properties and other 2D compounds, have been theoretically proposed and experimentally prepared \cite{Sili,silicine,beyond,phosp}. The Dirac semimetals with nonsymmorphic symmetry are a comparatively recent addition to the 2D material family. These materials feature Dirac points that are not gapped by spin-orbit coupling and are protected by nonsymmorphic lattice symmetry \cite{2D, bismuth}. Specific realizations of these materials have recently been proposed to occur in the nonsymmorphic monolayer film of $\alpha$-bismuthene, the monolayer of bismuth (MBi), black phosphorene, PtPb4, and in several others, and more details on the lattice structures and symmetries can be found in Refs. \cite{bismuth,PhysRevB.105.085101,C9NR00906J,doi:10.1126/sciadv.aar2317,He_2021,Liu2021,https://doi.org/10.1002/pssr.202200188,Wu2022}. 

In our recent paper (Ref.~{\cite{PhysRevB.105.085101}}), we  
have shown that nonsymmorphic Dirac semimetals are very interesting when the Zeeman coupling breaks the time-reversal symmetry and open a gap.  As a result, it creates sharp features in the optical absorption spectrum. In addition, these materials are intrinsically anisotropic and therefore exhibit the  phenomenon of optical birefringence. In the context of 2D materials, for a TRS broken system, the rotation of the polarization of the transmitted (reflected) light, i.e., the Faraday (Kerr) effect, can be used to deduce the off-diagonal element of the optical conductivity $\sigma_{xy}$ as was shown for monolayer graphene by Crassee et al. \cite{Crassee2011}. Nandkishore and Levitov have recently proposed that the quantum anomalous Hall state could be observed by measuring the Kerr rotation \cite{PhysRevLett.107.097402} in bilayer graphene samples. Inspired by the various interesting physics one can get from a gapped nonsymmorphic 2D Dirac semimetal, we investigate the magneto-optical properties like Faraday and Kerr rotation in our paper, owing to their invaluable scientific and engineering applications. In an applied context, these effects can be used for nondestructive material characterization, magneto-optical memory, magnetic field or current sensing, polarization rotators, and nonreciprocal optical devices (isolator, circulator)\cite{doi:10.1088/1468-6996/15/1/014602,Shoji_2016}.

\begin{figure}
    \centering
    \includegraphics[width=1.0\linewidth]{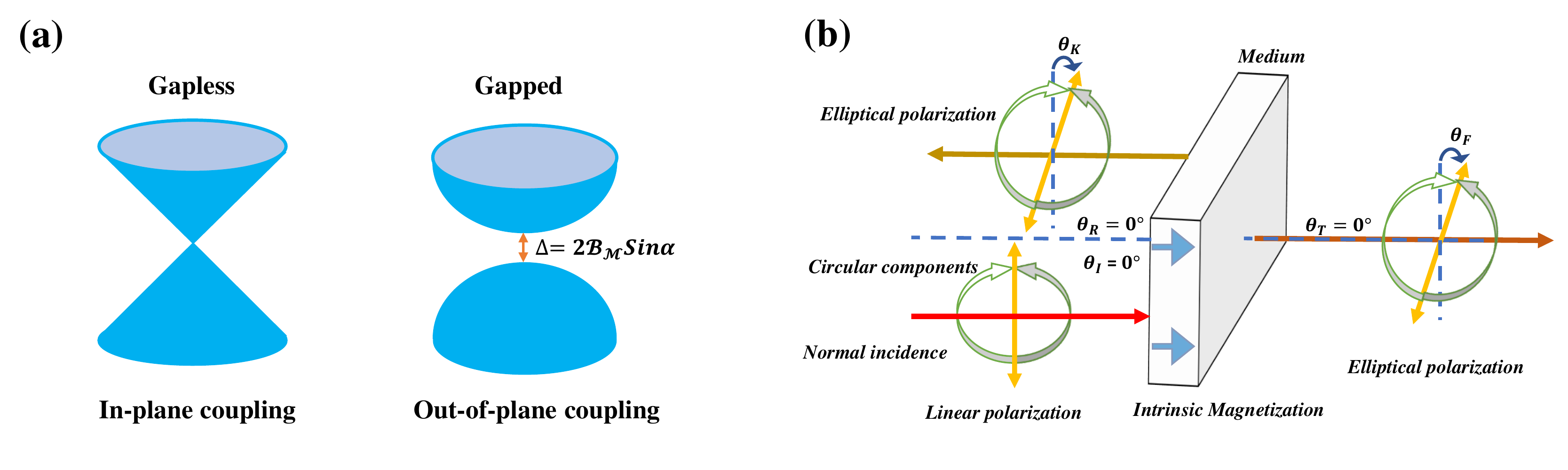}
\caption{\small \textbf{(a)} Schematic band diagram of a nonsymmorphic two-dimensional Dirac semimetal in the presence of a magnetic exchange potential. The exchange potential (in the form of a Zeeman term) creates the gap in otherwise linearly dispersing bands when it couples to out-of-plane spin. For an in-plane spin coupling, there is no gap. \textbf{(b)} Schematics of the Faraday and  Kerr rotation effects. In our study, we consider light  incident normal to the plane: thus, we keep the $\thetabold_{I}=0$ and $\thetabold_{R}=0$.} 
\end{figure}\label{fig:image0}

For a single atomic layer, whose thickness is much smaller than the wavelength of the incident light, the relationship between the Hall conductivity and Faraday (Kerr) angle $\theta_{F}(\theta_{K})$ can be derived by solving the Maxwell equations on the two sides of the atomic layer and matching solutions at the boundary. Such a derivation is worked out for bilayer graphene in Ref.~\cite{Crassee2011, https://doi.org/10.48550/arxiv.2008.02812},  for thin films of topological insulators in Ref.~\cite{PhysRevLett.105.057401, PhysRevB.82.161104, PhysRevB.84.205327}, and for thin films of topological Weyl semimetals in Ref.~\cite{Kargarian2015,S.nandy2}.  However, the chiral response of 2D nonsymmorphic Dirac materials has not been systematically investigated. 

Here we present our own derivation of the Faraday and Kerr rotation angles for a nonsymmorphic Dirac semimetal model, which includes intrinsic anisotropy. 
We consider an effective Zeeman coupling arising from exchange interactions with magnetic impurities or a nearby magnetized film~\cite{nano10091642,PhysRevB.101.045113,C9NR05698J}. The out-of-plane coupling of the exchange potential to the spin gives rise to a gap in the Dirac cone (proximity effect) whether an in-plane coupling does not\cite{PhysRevLett.105.057401,PhysRevB.95.115405,He_2021,Otrokov2019,doi:10.1073/pnas.2207681119}. This exchange potential can be tuned by changing the magnetization in the ferromagnetic insulator. Based on this model, we calculate the complete frequency-dependent optical conductivity tensor, including diagonal and off-diagonal (Hall) terms, and using that, we  construct the frequency-dependent transmission (reflection) matrix. When acting on the polarization vector of the incident light, this matrix gives out the polarization of the transmitted or reflected light. Through a simplified approach, we calculate the change in the direction of polarization of a linearly polarized wave upon transmission or reflection and show that the Faraday rotation angle is significantly enhanced as the frequency of the incident light approaches the threshold for optical absorption.  In the appendix, we provide complete analytical formulas to calculate the Faraday and the Kerr rotation angles for a general state of polarization of the incident light and show that the change in the state of polarization can be described by a single complex Faraday or Kerr rotation angle. As a special case,  we show that linearly polarized incident light acquires, upon transmission/reflection,  an elliptic component of  polarization, which is caused by the anisotropy of the system.

\section{Model Hamiltonian}

It is well known that two-dimensional materials with nonsymmorphic symmetries, such as glide mirror and screw-axis (\cite{PhysRevB.104.L201105,https://doi.org/10.48550/arxiv.2207.000201}), support symmetry-protected level crossings at time-reversal invariant points in the Brillouin zone. For example, $\alpha$-bismuthene ($\alpha$-Bi), with a glide-mirror symmetry $\tilde M_z$ \cite{bismuth,PhysRevB.105.085101} has symmetry-protected Dirac points at points $\bar X_1 = (\pi,0)$ and $\bar X_2=(0,\pi)$ of the Brillouin zone. The points $X_1$ and $X_2$ are not related to each other by a symmetry operation of the material. Therefore the corresponding Dirac cones will have, in general, different dispersions and different energies at the crossing point. 
We note that having symmetry-protected Dirac points does not guarantee that these points will occur in close vicinity of the Fermi level, where they can control the electronic properties of the material.  We will assume that at least one of the Dirac points is indeed at the Fermi level: this is not too far from reality for the $X_1$ point of $\alpha$-Bi.

A general Hamiltonian that captures the low-energy dispersion of nonsymmorphic Dirac semimetals near the Dirac point has been derived in Ref. (\cite{bismuth, PhysRevB.105.085101}). It has the form 
\begin{equation}\label{Hamiltonian}
      H = \rho v k_x(\cos\alpha \ \sigma_x\otimes\tau_z +\sin\alpha \ \sigma_0\otimes\tau_y)+v k_y\ (\sigma_y\otimes\tau_z)\, ,
        \end{equation}
where the Pauli matrices $\sigma$ and $\tau$ refer to spin degrees of freedom and orbital degrees of freedom, respectively -- the orbital degrees of freedom are associated with $p$-orbitals of the atomic sites of the $\alpha$-Bismuthene lattice \cite{bismuth}. The essential glide-mirror symmetry is represented by the operator $\tilde M_z= \sigma_z\otimes\tau_y$, while the remaining time-reversal, inversion, and the $x$-mirror symmetries are represented by $T =-i\sigma_y\otimes\tau_0 K$, $P = \sigma_0\otimes\tau_x$ and $M_x =-i \sigma_x\otimes\tau_x$ respectively.

In Eq.~(\ref{Hamiltonian}) we have introduced $\rho$ as the anisotropy factor $\rho = \frac{v_x}{v_y}$ which gives the ratio of the Fermi velocities along $x$ and $y$-direction, where $v_y=v$ and $\rho v = v_x$. For example, in $\alpha$-Bi, the Dirac cones at $X_1$ and $X_2$ have  $\rho = 1.86$ and $\rho = 0.25$, respectively (again, this difference is allowed because the two valleys at $X_1$ and $X_2$ are not connected by any crystal symmetries).  The angle $\alpha$, which we call the "mixing angle", is also an intrinsic parameter of the model and is related to the spin-orbit coupling of the system. While $\alpha$ does not affect the dispersion of the bands in the absence of a gap, it plays a central role in determining the magnitude of the gap that appears when the time-reversal symmetry is broken, and a gap opens up.

The $\tilde M_z$ symmetry allows us to decompose the $4 \times 4$ Hamiltonian into two 2$\times$2 blocks representing the $\tilde M_z$ even sector (eigenvalue +1) and odd sector (eigenvalue -1). Since the eigenvalues are the same in the two sectors, for convenience, we choose to work within the even sector, corresponding to $\sigma_z =1$, $\tau_y=1$ and $\sigma_z=-1$, $\tau_y=-1$. Choosing a convenient basis as described in Ref.~\cite{PhysRevB.105.085101}) $\sigma_z$ becomes $\sigma_x$, $\sigma_x$ becomes $\sigma_y$, and $\sigma_y$ becomes $\sigma_z$. The reduced  Hamiltonian is written as
\begin{equation}\label{reduced_ham}
    \tilde H_{\tilde M_z=1}= \rho v k_x (\sigma_x \cos \alpha  \ +  \sigma_y \sin \alpha) + v k_y\sigma_z\,,
\end{equation}
whose eigenvalues are
\begin{equation}
    E=\pm v\sqrt{\rho^2k_x^2+k_y^2}
\end{equation}
Notice that we could reduce the whole Hamiltonian to 2$\times$2 form only because the system has the glide mirror symmetry $\tilde M_z$.

A magnetic exchange potential coupling to the out-of-plane spin can arise from a magnetic proximity effect and takes the form of a Zeeman term  $\Bc_{\Mc} \sigma_{z} \tau_0$ in the Hamiltonian, where $\Bc_{\Mc}$ is the effective out-of-plane  field arising from the exchange coupling. The glide mirror symmetry is  preserved in the presence of this term. We can, therefore, still use the reduced Hamiltonian Eq.~(\ref{reduced_ham}) under the glide mirror symmetry operator. To this end, we decouple this term for even and odd sectors (the same as we do for the Hamiltonian). We note that $\Bc_{\Mc} \tau_0 \sigma_z$ can be reduced to $\Bc_{\Mc} \sigma_z$ in the even sector and $-\Bc_{\Mc} \sigma_z$ in the odd sector. Continuing to work in the even sector with the modified basis and also with the term added, we find

\begin{equation}\label{Ham_z}
   H'_{M_z=1}=(\rho v k_x \cos \alpha+\Bc_{\Mc} )\sigma_x\ + \ \rho v k_x \sin \alpha \sigma_y\ +\ v k_y \sigma_z \,,
\end{equation}
whose eigenvalues are 
    \begin{equation}\label{eig_z}
  E=\pm v\sqrt{\left( \rho k_x + \frac{\Bc_{\Mc}}{v} \cos\alpha \right)^2+k^2_y+\frac{{\Bc^2_{\Mc}}}{v^2} \sin^2\alpha}\,.
\end{equation}
\begin{figure}
    \centering
    \includegraphics[width=1.0\textwidth]{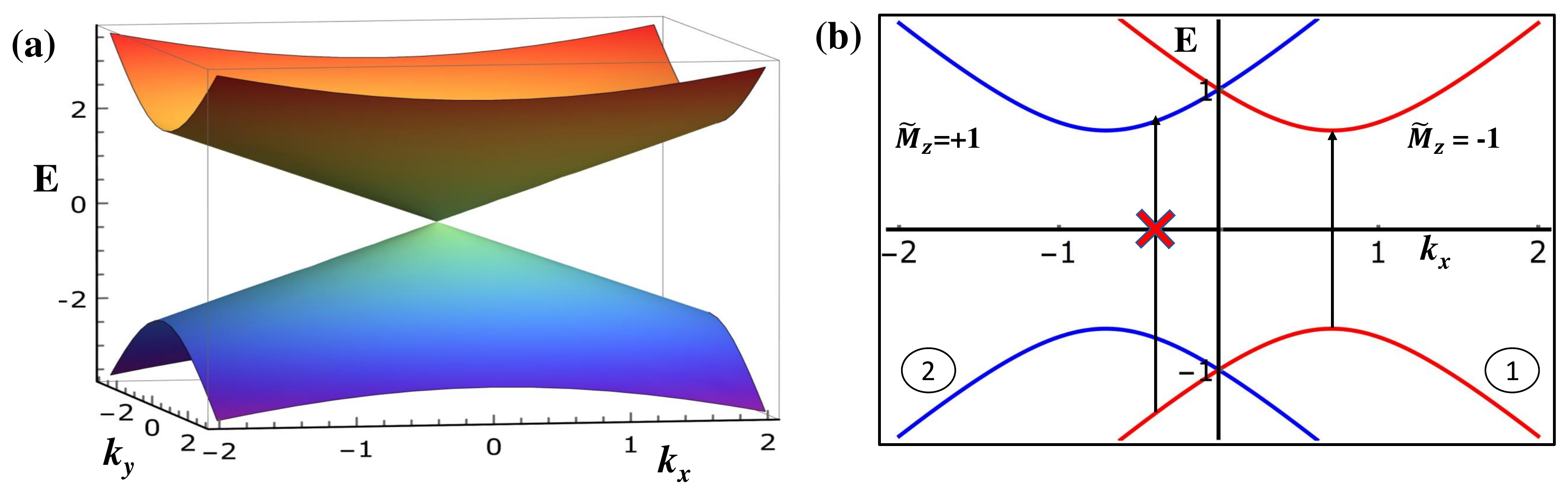}
\caption{\small \textbf{(a)} The band dispersion of $\alpha$-Bi near the $X_1$ Dirac point.\textbf{(b)} This figure shows that only intra-valley transitions are possible as the two valleys hold opposite eigenvalues for $\tilde M_z$ when there is a gap present in the system.}
\label{fig:image1}
\end{figure}
The main qualitative effect of the exchange potential is the appearance of a gap in the excitation spectrum. We find the band gap using Eq.~(\ref{eig_z}) as:
\be\label{Gap1}
\Delta = 2\min |\bar E|=2 \Bc_{\Mc}\sin\alpha ,
\ee
In practice, this gap is expected to be on the order of a few meV \cite{nano10091642,PhysRevB.101.045113,C9NR05698J,PhysRevB.95.115405,He_2021,Otrokov2019,doi:10.1073/pnas.2207681119}. As anticipated, Eq.~(\ref{Gap1}) offers a direct way to determine the mixing angle of our model from optical absorption experiments. 
The eigenfunctions for this Hamiltonian are
\begin{equation}
    \ket{\psi_{c\kv}}=
\begin{pmatrix}
\cos\frac{\theta}{2}\ e^{-i\gamma}\\
\sin\frac{\theta}{2}\\
\end{pmatrix};~~
\ket{\psi_{v\kv}}=
\begin{pmatrix}
\sin\frac{\theta}{2}\ e^{-i\gamma}\\
-\cos\frac{\theta}{2}
\end{pmatrix},
\end{equation}
with 
\begin{equation}
   \tan\gamma=\frac{\rho k_x \sin\alpha}{\rho k_x \cos\alpha+\frac{\Bc_{\Mc}}{v}};~~
   \tan\theta=\frac{\sqrt{(\rho k_x +\frac{\Bc_{\Mc}}{v}\cos\alpha)^2+\frac{\Bc^2_{\Mc}}{v^2} \sin^2\alpha}}{k_y}\,,
\end{equation}
which clearly depends on the effective Zeeman field $\Bc_{\Mc}$. 
To proceed with the study of optical properties, we need the current operators $\hat j_x$ and $\hat j_y$, which are obtained by taking the derivative of our Hamiltonian (Eq.~(\ref{Ham_z})) with respect to $k_x$ and $k_y$ respectively,   i.e., ${\hat \jv} =\frac{\partial H}{\partial \kv}$ with components
\begin{equation}\label{CurrentOperators}
 \hat j_x=\rho v \left(\cos\alpha \sigma_x+\sin\alpha \sigma_y\right)\,,~~~~\hat j_y=v \sigma_z\,.
\end{equation}
These operators commute with the glide mirror operation $\tilde M_z$, which makes the transitions preserve the parity of $\tilde M_z$. Fig.~\ref{fig:image1}(b) shows that the two valleys have opposite parity under the glide mirror symmetry; hence, inter-valley transitions are not allowed.

\section{The transmission/reflection matrix}
To treat effects like Faraday (Kerr) rotation for a single atomic layer, we need to compute the $2 \times 2$ matrices that connect the polarization state of the incoming light (a 2-dimensional complex vector in the plane perpendicular to the direction of propagation of the light) to the polarization states of the transmitted and reflected light. (We assume here for simplicity that the incident, transmitted, and reflected light all travel along the $z$-axis perpendicular to the layer).  In the standard treatment, these matrices are obtained by matching the electric fields on opposite sides of the layer, taking into account the two-dimensional electronic current that flows in the layer under the action of the electric field, causing  the magnetic field to change discontinuously across the layer. In what follows, we will assume that the layer is much thinner than the wavelength of the light so that the thickness of the layer can be neglected, and positions immediately before and after the layer will be labeled by $z=0^-$ and $z=0^+$, respectively.

 We  work in the gauge where the scalar potential and the $z$ component of the vector potential of the incoming electromagnetic wave vanish.  The transverse components of the vector potential $\Av=(A_x,A_y)$  satisfy the Maxwell equation (in Gaussian units)
\begin{equation}\label{maxwell}
    -\partial_{z}^2 \Av(z) -\frac{\omega^2}{c^2}\Av(z)=\frac{4\pi}{c}\Jv(z)\,,
\end{equation}
where $\omega$ is the wave's frequency  and $\Jv(z)$ is the three-dimensional current density, concentrated in an infinitesimal region around $z=0$.  We define the two-dimensional current density in the layer as 
\begin{equation}
 \jv=\int dz \Jv(z)\,,
\end{equation}
where the integral over $z$, while formally extending to infinity, converges within a small vanishing region corresponding to the thickness of the layer. 
The two-dimensional current density $\jv$ is related to the electric field by the frequency-dependent electrical conductivity tensor $\sigmabold(\omega)$:
\begin{equation}
 \jv=\sigmabold(\omega)\cdot \Ev\,.
\end{equation}
The electric and magnetic field are given respectively by $\Ev = \frac{i \omega}{c}\Av$ and $\Hc = \zv \times \Av'$ where $\Av'$ is the derivative of $\Av$ with respect to $z$.

Integrating Eq.~(\ref{maxwell})  over z from $z=-\epsilon$ to $z = +\epsilon$, with $\epsilon$ tending to zero we obtain:
\begin{equation}\label{BC1}
   \delta\Av'\equiv  -\Av'|^{\epsilon}_{-\epsilon}=\frac{4\pi}{c}\jv=\sigmabold\cdot \Ev\,.
\end{equation}
This tells us that $\Hc$  ``jumps" across the layer.
At the same time, the vector potential itself (and hence the electric field) is continuous across the layer, i.e., 
\begin{equation}\label{BC2}
   \delta\Av \equiv \Av(\epsilon)-\Av(-\epsilon)=0\,.
\end{equation}
We seek a solution to the form
\begin{equation}\label{vect_pot}
\Av (z) \propto \begin{cases}
         \ev_{i}e^{ikz}+{\bf R}\cdot \ev_{i}e^{-ikz},&  z < 0\\
            {\bf T}\cdot\ev_{i}e^{ikz}\,,      & z > 0
            \end{cases}
\end{equation}
where $\ev_{i}$ is a two-dimensional complex vector  describing the state of polarization of the incident  polarization of the incoming wave in the $(x,y)$ plane, $\Rv$ and $\Tv$ are the reflection, and the transmission matrices  -- $2\times 2$ complex matrices acting on the polarization.  Imposing the boundary conditions~(\ref{BC1}) and ~(\ref{BC2}) we find 
\begin{equation}\label{TransmissionMatrix}
     \Tv =\left(1+a.\sigmabold \right)^{-1}\,,
\end{equation}
\begin{equation}\label{ReflectionMatrix}
 \Rv =\Tv-{\bf 1}=\-a\sigmabold\cdot \left(1+a.\sigmabold \right)^{-1}\,,
\end{equation}
where we have defined $a\equiv\frac{2\pi}{c}$. As previously mentioned, the transmission matrix (\Tv), acting on the incoming polarization vector ($\ev_{i}$), gives the polarization of the transmitted light $\ev_{t}$ and similarly, the reflection matrix (\Rv) gives the polarization of the reflected light $\ev_{r}$.
\begin{equation}\label{optical response funct}
    \ev_{t}=\Tv\cdot \ev_{i};~~\ev_{r}=\Rv\cdot\ev_{i}
\end{equation}
Now that we have established the relation between the transmission/reflection matrices and the conductivity tensor, we will calculate the latter.

\section{Calculation of the conductivity tensor}
In this section, we calculate the conductivity tensor using linear response theory, as described in Ref.~(\cite{giuliani_vignale_2005}). The conductivity tensor for our model system is given by
\begin{equation}\label{cond_tensor2}
    \sigma_{\alpha \beta}(\omega) = \frac{i e^2}{\omega}\chi_{j_{\alpha} j_{\beta}}(\omega)\,,
\end{equation}
where $\chi_{j_{\alpha} j_{\beta}}$ is the current-current response function
\begin{equation}\label{Current-CurrentResponse}
\chi_{j_{\alpha} j_{\beta}}(\omega)=\sum_{\kv}\left\{\frac{\langle\psi_{c\kv}|\hat j_{\alpha}|\psi_{v\kv}\rangle\langle\psi_{v\kv}|\hat j_{\beta}|\psi_{c\kv }\rangle}{\omega-\omega_{cv}(\kv)+i\eta}- \frac{\langle\psi_{c\kv}|\hat j_{\beta}|\psi_{v\kv}\rangle\langle\psi_{v\kv}|\hat j_{\alpha}|\psi_{c\kv}\rangle}{\omega+\omega_{cv}(\kv)+i\eta}\right\}\,,
\end{equation}
where $\psi_{c\kv}$ and $\psi_{v\kv}$ are the Bloch states of the conduction and  valence band, respectively, and $\omega_{cv}(\kv)$ is the difference of their energies. $\eta$ is an infinitesimal positive real number.  The operators $\hat j_\alpha$ are defined in Eq.~(\ref{CurrentOperators}).  We notice that the ``diamagnetic'' contribution to the conductivity~\cite{giuliani_vignale_2005} is absent because our model Hamiltonian is linear in $\kv$.

In the next two subsections, we treat this tensor's off-diagonal and diagonal components separately.

\subsection{Off-diagonal component of the conductivity tensor ($\sigma_{xy}$)}
We calculate the off-diagonal component of the conductivity tensor first. We separately consider the  imaginary and the real parts of the response function $\chi_{j_{x} j_{y}}$.  For the imaginary part, we find (in the limit $\eta \to 0$)
\begin{equation}
 \chi''_{j_{x}j_{y}}=2\rho v^2 \sum_{k}\frac{2\omega}{\omega^2-\omega_{cv}^2(\kv)} \frac{\Bc_{\Mc}\sin\alpha}{E(\kv)}\,,
\end{equation}
where the integral over $\kv$ is done according to the Cauchy principal value prescription. 
This is different from zero both below ($|\omega| \le \Delta$) and above the gap ($|\omega| \ge \Delta$).  For the real part of $\chi_{j_{x} j_{y}}$ we find
\begin{equation}
 \chi'_{j_{x}j_{y}} =2\pi \rho v^2 \sum_{\kv} \frac{\Bc_{\Mc}\sin\alpha}{E(\kv)}\left[ \delta(\omega-\omega_{cv}(\kv)) + \delta(\omega+\omega_{cv}(\kv))\right]\,,
\end{equation}
which differs from zero only above the gap, i.e., for $|\omega|>\Delta$. In writing these expressions, we have included a factor $2$ arising from the double degeneracy of the energy eigenvalues associated with glide-mirror parities $\tilde M_z=1$ and $\tilde M_z=-1$ (see discussion before  Eq.~(\ref{Ham_z})). Then, making use of Eq.~(\ref{cond_tensor2}) we obtain the real part of the off-diagonal conductivity 
\begin{equation}\label{hall_con1}
         \sigma'_{xy}(\omega) =-\frac{ e^2}{2\pi\hbar}\frac{\Delta}{\omega}\log\abs{\frac{\Delta-\omega}{\Delta+\omega}}\ (\text{Here reinstating the $\hbar$ to match dimension})\,.
\end{equation}
This expression shows that the Hall conductivity has logarithmic divergence when $\omega$ approaches the gap frequency $\Delta$. 
Similarly, the imaginary part of the off-diagonal conductivity is given by
 \begin{equation}
    \begin{split}
        \sigma''_{x y} =\frac{e^2}{4\hbar} \frac{\Delta}{\omega} \Theta (|\omega|-\Delta)\,,
    \end{split}
\end{equation}  
where $\Theta(x)$ is the Heaviside step function $\Theta(x)=1$ for $x>0$ and $\Theta(x)=0$ for $x<0$.  Notice that $\sigma_{xy}$ vanishes for $\Bc_{\Mc}=0$,  leaving the system with only diagonal components of the conductivity in the absence of $\Bc_{\Mc}$.
\begin{figure}[t]
    \centering
    \includegraphics[width=1.0
\textwidth]{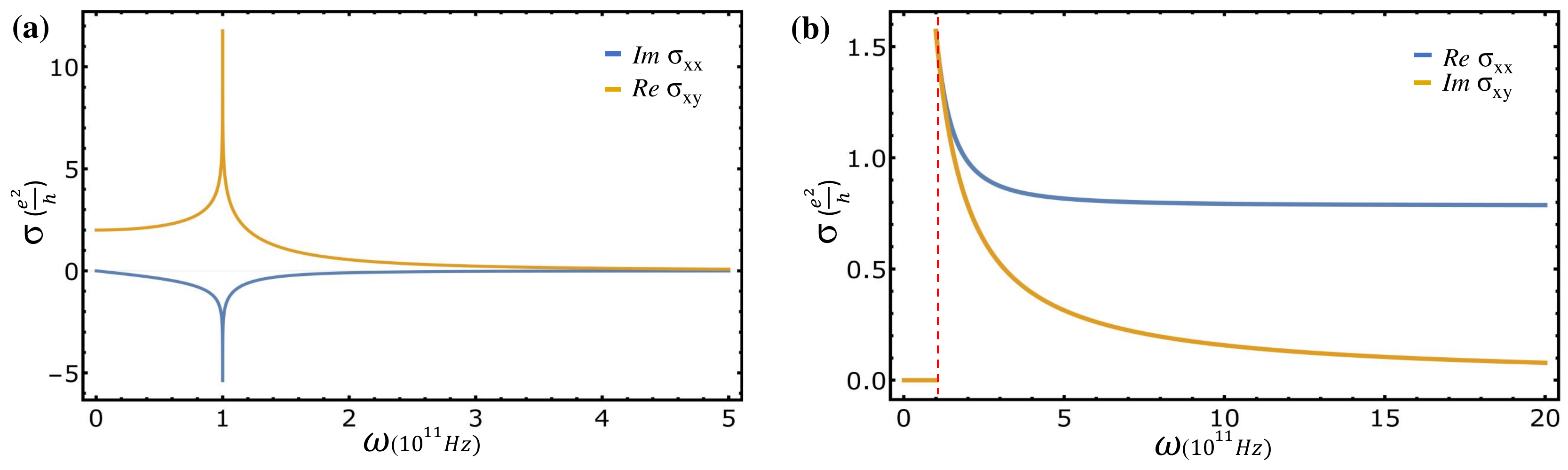}
\caption{\small \textbf{(a)} Plots of $\sigma'_{xy}$ and $\sigma''_{xx}$ vs $\omega$.    Both exhibit a logarithmic singularity at the absorption edge $\omega=\Delta$ ($\Delta=2\Bc_{\Mc}\sin\alpha=1$ in this plot); however, the divergence in $\sigma'_{xy}$ has a stronger prefactor.   Notice that $\sigma'_{xy}$ has a finite value ($2\frac{e^2}{h}$) at zero frequency, whereas $\sigma''_{xx}$ vanishes.  \textbf{(b)} Plots of $\sigma''_{xy}$ and $\sigma'_{xx}$ vs $\omega$.  These functions vanish for frequencies below the gap. The red dashed line marks the absorption edge.}
\label{fig:image2}
\end{figure}
\subsection{Diagonal components of the conductivity tensor ($\sigma_{xx},\sigma_{yy}$)}
Following the same procedures as in the previous section, we obtain the following formulas for the  real and imaginary parts of the diagonal response functions $\chi_{j_{x} j_{x}}$  and $\chi_{j_{y} j_{y}}$:  
\begin{equation}
    \chi'_{j_{x} j_{x}}(\omega) =2\rho^2 v^2 \sum_{\kv}\frac{2\omega_{cv}(\kv)}{\omega^2-\omega_{cv}^2(\kv)}\left(\frac{k^2\sin^2\phi(\kv)+\Bc^2_{\Mc}\sin^2\alpha}{E^2(\kv)}\right);~~~~\chi'_{j_{y} j_{y}}(\omega)=\frac{\chi'_{j_{x} j_{x}}(\omega)}{\rho^2} \ .
\end{equation}
and
\begin{equation}
\chi''_{j_{x} j_{x}}(\omega)=-2\pi\rho^2 v^2 \sum_{\kv}\left[\delta(\omega-\omega_{cv}(\kv))-\delta(\omega+\omega_{cv}(\kv)) \right]\left(\frac{k^2\sin^2\phi(\kv)+\Bc^2_{\Mc}\sin^2\alpha}{E^2(\kv)}\right)\,.
\end{equation}
where $\phi(\kv)=\arctan\left(\frac{k_{y}}{\rho k_x +\frac{\Bc_{\Mc}}{v}\cos\alpha}\right)$ and $\chi''_{j_{y} j_{y}}=\chi''_{j_{x} j_{x}}/\rho^2$. Using Eq.~(\ref{cond_tensor2}), we can conclude that the imaginary and real parts of the conductivity are given by
\begin{equation}
        \sigma''_{xx}(\omega)=\frac{\rho e^2}{4\pi\hbar} \left[\frac{\Delta}{\omega}+ \frac{(\Delta^2+\omega^2)}{2\omega^2}\ \log\abs{\frac{\Delta-\omega}{\Delta+\omega}} \right]\ ;~~\sigma''_{yy}(\omega)=\frac{\sigma''_{xx}(\omega)}{\rho^2}\,,
\end{equation}
and
\begin{equation}
            \sigma'_{xx}(\omega)=\frac{\rho e^2}{8\hbar}\left[1+\frac{\Delta^2}{\omega^2}\right] \Theta(|\omega|-\Delta);~~~\sigma'_{yy}(\omega)=\frac{\sigma'_{xx}(\omega)}{\rho^2}\ .
\end{equation}
The anisotropy parameter $\rho$ is the ratio between the velocities in the $x$ and $y$-directions and reduces to $1$ in the isotropic case.

\section{Calculation of the Faraday and Kerr rotation angles}
From Eq.~(\ref{TransmissionMatrix}) we see that the transmission matrix is given by
\be\label{TransmissionMatrix2}
{\bf T} = \frac{1}{N} \left(\begin{array}{cc} 1+a\sigma_{yy}&-a\sigma_{xy}\\-a\sigma_{yx}&1+a\sigma_{xx}\end{array}\right)\,,
\ee
where $\sigma_{xy}=-\sigma_{yx}$ and  
\begin{equation}
 N= (1+a\sigma_{xx})(1+a\sigma_{yy})+(a\sigma_{xy})^2
\end{equation}
On the other hand, from Eq.~(\ref{ReflectionMatrix}),  
we find that the components of the reflection matrix are given by
\be\label{ReflectionMatrix2}
{\bf R} =  \left(\begin{array}{cc} T_{xx}-1& T_{xy}\\T_{yx}&T_{yy}-1\end{array}\right)\,,
\ee
From these formulas, we now proceed to extract the Faraday and Kerr rotation angles.
\subsection{Isotropic system} 
For the isotropic system we have $\sigma_{xx}=\sigma_{yy}$ ($\rho=1$) which leads to $T_{xx}=T_{yy}$. In this case, it is easy to see that the states of circular polarization are eigenstates of the transmission and reflection matrix, meaning that these states of polarization are not changed upon transmission or reflection but suffer a phase shift.

\begin{figure}[t!]
    \centering
    \includegraphics[width=1.0\textwidth]{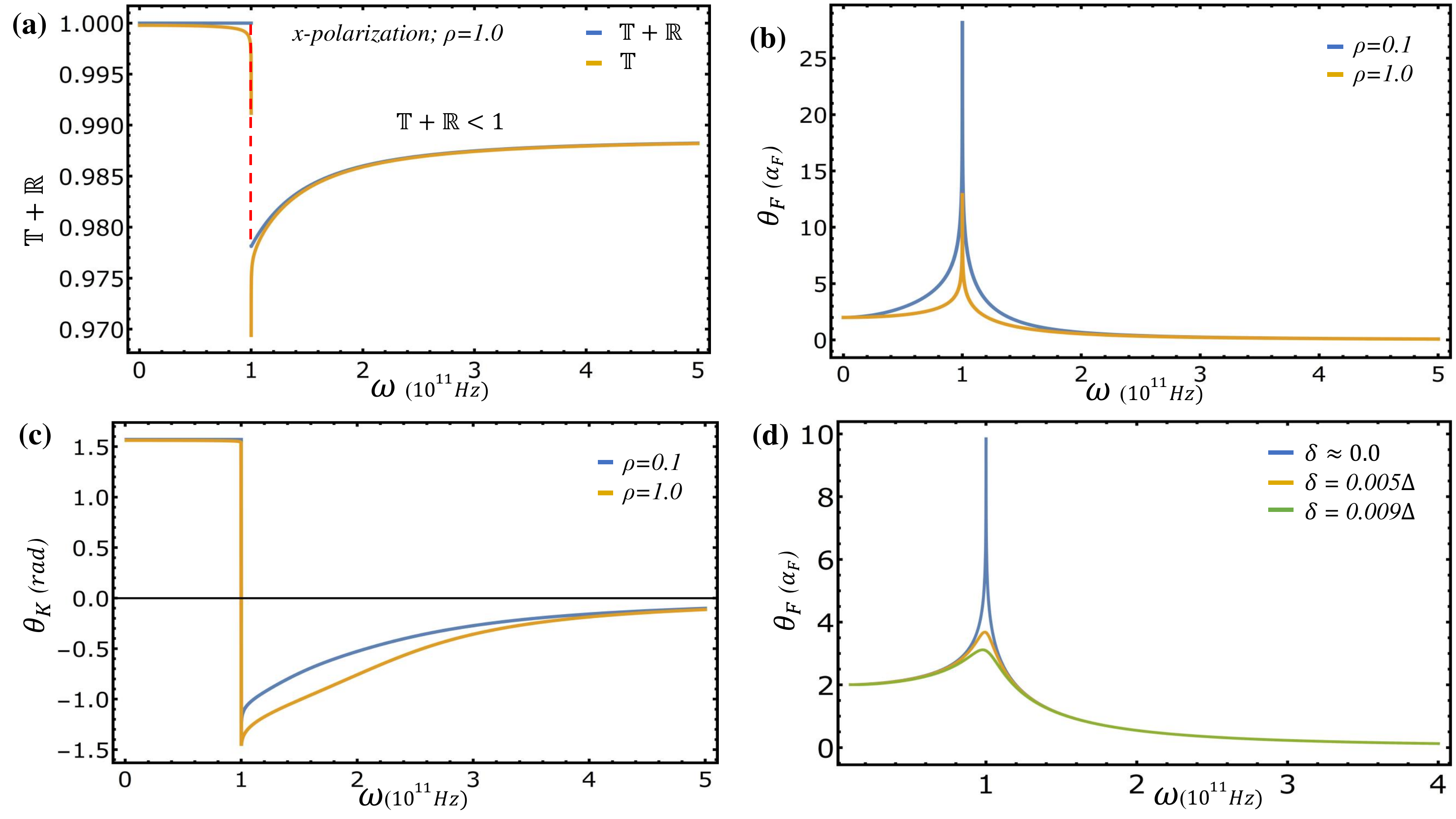}
    \caption{\small \textbf{(a)} Plot of transmittance and transmittance+reflectance vs frequency in units of $\Delta=10^{11}$ Hz. Notice that transmittance and reflectance add to $1$ below the absorption edge, marked by the vertical red dashed line. The value of the reflectance can be inferred from the small difference between these two plots.  Above the absorption edge, the sum $\mathbb{T}+\mathbb{R}$ is less than $1$ due to the optical absorption. \textbf{(b)} Faraday rotation angle vs frequency.  The  orange line is for the isotropic case ($\rho=1.0$),  and the blue line is for the  anisotropic case ($\rho=0.1$) calculated as described in Section VB. I Notice that the same zero-frequency value ($\alpha_{F}$) is independent of the anisotropy parameter $\rho$. The enhancement of the Faraday rotation angle at the absorption threshold is much larger in the anisotropic case. \textbf{(c)} Kerr rotation angle vs frequency. The orange and blue lines are for $\rho=1$ and $\rho=0.1$, respectively.   \textbf{(d)} The effect of disorder broadening ($\delta=1/\tau$) on the  Faraday rotation angle, calculated by the procedure described in Section VI.}
    \label{fig:image3}
\end{figure}

The Faraday and Kerr rotation angles are defined as half of the difference between the phase shifts of the  left circular polarized (LCP) wave and the right circular polarized (RCP) waves in the transmitted and reflected wave, respectively \cite{https://doi.org/10.48550/arxiv.2008.02812,PhysRevB.84.235410,PhysRevLett.105.057401}: 
\begin{equation}\label{chiral_faraday_def}
    \theta_F=\frac{1}{2} Arg \left[\frac{E^{t}_{LCP}}{E^{t}_{RCP}}\right] ;~~~~~\theta_K=\frac{1}{2}Arg \left[\frac{E^{r}_{LCP}}{E^{r}_{RCP}}\right] .
\end{equation}
The $x$ and $ y$ components of the electric field are related to the LCP and RCP components as follows: 
\begin{equation}\label{chiral_iso}
    \begin{pmatrix}
        E_{x}\\
        E_{y}\\
    \end{pmatrix}=\frac{1}{\sqrt{2}}\begin{pmatrix}
        1 & 1\\
        -i & i\\
    \end{pmatrix} \begin{pmatrix}
        E_{LCP}\\
        E_{RCP}\\
    \end{pmatrix}
\end{equation}
Substituting this in the equations that relate the transmitted and reflected fields to the incident field\footnote{Considering we have the incident wave linearly polarized along the $x$-direction.}, namely, 
\begin{equation}\label{Trans_Reflect_iso}
            \begin{pmatrix}
            E^{t}_{x}\\
            E^{t}_{y}\\
        \end{pmatrix}= \begin{pmatrix}
            T_{xx}\\ 
            T_{yx}\\
        \end{pmatrix} E^{i}_{x};~~
        \begin{pmatrix}
            E^{r}_{x}\\
            E^{r}_{y}\\
        \end{pmatrix} = \begin{pmatrix}
            R_{xx}\\ 
            R_{yx}\\
        \end{pmatrix} E^{i}_{x} \ .
\end{equation}
we easily find
\begin{equation}\label{theta_chiral2}
    \theta_F=\frac{1}{2} Arg \left[\frac{T_{xx}+ iT_{xy}}{T_{xx}- iT_{xy}}\right]=\frac{1}{2} Arg \left[\frac{1+ a\sigma_{xx}+i a \sigma_{xy}}{1+ a\sigma_{xx}-i a \sigma_{xy}}\right] \ ,
\end{equation}
and
\begin{equation}\label{kerr_chiral}
    \theta_{K}=\frac{1}{2} Arg\left[\frac{R_{xx}+ iR_{xy}}{R_{xx}-iR_{xy}}   \right]=\frac{1}{2} Arg\left[\frac{\sigma_{xx}- i\sigma_{xy}}{\sigma_{xx}+i\sigma_{xy}} \frac{1+ a(\sigma_{xx}+i \sigma_{xy})}{1+ a(\sigma_{xx}-i \sigma_{xy})} \right]\,.
\end{equation}
At zero frequency, the Hall conductivity has a nonzero value ($\sigma_{xy}(0)=2\frac{e^2}{h}$), but the longitudinal one is zero. This results in a nonzero  Faraday rotation angle at zero frequency:
\begin{equation}
    \theta_F(0)=\frac{1}{2} Arg \left[\frac{1-i a \sigma_{xy}(0)}{1+i a \sigma_{xy}(0)}\right]=\frac{1}{2} Arg \left[\frac{1-2 i \alpha_{F}}{1+ 2 i \alpha_{F}}\right]=2\alpha_{F}\,, 
\end{equation}
where $\alpha_{F}=\frac{e^2}{\hbar c}$ is the fine structure constant. Plotting the Faraday angle as a function of frequency, we see that, similar to the conductivities, it has a sharp rise at the absorption edge. Still, instead of diverging, it reaches a maximum theoretical   value  $\frac{\pi}{2}$. More precisely, when we choose the frequency very near to the absorption edge, we find that
\begin{equation}
    \theta_F(\omega\sim \Delta) \approx \frac{1}{2} Arg \left[\frac{a\sigma_{yy}-i a \sigma_{xy}}{a\sigma_{yy}+i a \sigma_{xy}}\right] \approx \frac{1}{2} Arg \left[\frac{-i a \sigma_{xy}}{+i a \sigma_{xy}}\right]=\frac{\pi}{2}\,, 
\end{equation}
where the second approximate equality follows from the fact that the logarithmic divergence of $\sigma_{xy}(\omega)$ for $\omega \to \Delta$  has a larger prefactor compared to the logarithmic divergence of $\sigma_{xx}(\omega)$.  
The range of frequencies around $\omega=\Delta$  in which the Faraday angle would be close to the theoretical limit $\frac{\pi}{2}$ is exponentially small  ($\simeq e^{-1/\alpha_{F}}$) and unobservable in practice.   Nevertheless,  an order of magnitude increase in $\theta_F$ is visible in Fig.~\ref{fig:image3}(b).  While the peak value of $\theta_F$ is a far cry from $\pi/2$, it is still more significant than any value previously measured near the absorption edge for two-dimensional systems \cite{PhysRevB.101.045426, PhysRevB.84.235410}. 

A similar analysis can be done  for the Kerr rotation angle. 
From  Eq.~(\ref{kerr_chiral}), we see that in the low-frequency regime, the Kerr rotation angle is  $\theta_{K} \approx \frac{\pi}{2}$.  This large value describes that left and right circularly polarized waves are reflected with amplitudes of opposite signs, corresponding to a phase difference of $\pi$.  In Fig.~\ref{fig:image3}(c), it can be seen that this value remains constant below the absorption edge, but it reverses its sign as the frequency crosses the absorption edge.

As a final point, in Fig.~\ref{fig:image3}(a) we plot the transmittance $\mathbb{T}$ and the reflectance $ \mathbb{R}$ defined as follows\cite{PhysRevB.84.235410}: 
\begin{equation}
    \mathbb{T}= \frac{1}{2} \left(\abs{T_{LCP}}^2+\abs{T_{RCP}}^2\right);~~~  \mathbb{R}= \frac{1}{2} \left(\abs{R_{LCP}}^2+\abs{R_{RCP}}^2\right) .
\end{equation}
Below the absorption edge, we verify that we have $\mathbb{T}+\mathbb{R}=1$, in accordance with the fact that there is no absorption of energy in this frequency range.  Above the absorption edge, on the other hand, we find  $\mathbb{T}+\mathbb{R}<1$, where the difference refers to the existence of absorption above the gap. In our recent study (Ref.~\cite{PhysRevB.105.085101}), we have shown when there is a gap present in this system, it will give rise to significant absorption above the gap, and this is just another confirmation of the same.

\subsection{Anisotropic system}
The calculation of the Faraday angle for an anisotropic system presents us with some basic difficulties.  The right and left circularly polarized states are no longer eigenstates of the transmission and reflection matrix: rather, an incident LCP wave will acquire an RCP component upon transmission/reflection, and similarly, an incident RCP wave will acquire an LCP component.  If, on the other hand, we start with a linearly polarized incident wave, then the transmitted/reflected waves will not be in a state of linear polarization, having acquired elliptic components. It is still possible to define a rotation angle, as discussed in Appendix A. Still, this rotation angle will then depend on the orientation of the incident linear polarization with respect to the crystallographic axes.

In this section, we adopt a practical definition of the Faraday and Kerr rotation angles, which avoids these difficulties and yields values independent of the state of polarization of the incident wave.  The idea is to generalize the definition we used in the isotropic case and define the magneto-optical rotation angles as half the difference of the phase shifts associated with the exact eigenstates of the transmission and reflection matrices. 

To do this, we will diagonalize the transmission matrix first.   In the chiral basis \footnote{The chiral basis states are related to the $x-y$ basis states by the unitary transformation $U=\frac{1}{\sqrt{2}}\left(\begin{array}{cc}1&1\\i&-i\end{array}\right)$, $U^\dagger=\frac{1}{\sqrt{2}}\left(\begin{array}{cc}1&-i\\1& i\end{array}\right)$. We then have $\tilde v = U^\dagger v$ where $v$ is the representative of a state in the $x-y$ basis and $\tilde v$ is the representative of the same state in the chiral basis.} this matrix has the following structure
\be\label{Chiral_trans}
    \tilde{\Tv}=U^{\dagger}\cdot \begin{pmatrix}
T_{xx} & T_{xy}\\
-T_{xy} & T_{yy}\\
\end{pmatrix}\cdot U=
\begin{pmatrix}
\bar T+iT_{xy}  & \Delta T\\
 \Delta T &\bar T -iT_{xy} \\
\end{pmatrix}
\ee
where
\be
 \bar T \equiv \frac{T_{xx}+T_{yy}}{2}\,,~~~\Delta T \equiv \frac{T_{xx}-T_{yy}}{2}\,.
\ee
Its eigenvalues can be written as 
\begin{equation}\label{chiraltran_eig}
    T_{\pm}=\bar T \pm i\sqrt{T_{xy}^2-\Delta T^2};
\end{equation}
Similarly, the eigenvalues of the reflection matrix are given by 
\begin{equation}\label{chiralreflect_eig}
    R_{\pm}=\bar R \pm i\sqrt{R_{xy}^2-\Delta R^2};
\end{equation}
In terms of $T_+$ and $T_-$, the Faraday rotation angle is expressed as
\begin{equation}\label{faraday_prag}
    \theta_F\equiv\frac{1}{2} Arg \left[\frac{T_{+}}{T_{-}}\right]=\frac{1}{2} Arg \left[\frac{\bar T + i\sqrt{T_{xy}^2-\Delta T^2}}{\bar T - i\sqrt{T_{xy}^2-\Delta T^2}}\right]\ ,
\end{equation}
In terms of the conductivities, this becomes
\begin{equation}
    \theta_{F}= \frac{1}{2} Arg\left[ \frac{2+a\tilde\sigma_{xx}\left(\rho +\rho^{-1}\right)+i a \sqrt{4\sigma^2_{xy}-\tilde\sigma^2_{xx}\left(\rho -\rho^{-1}\right)^2}}{2+a\tilde\sigma_{xx}\left(\rho +\rho^{-1}\right)-i a \sqrt{4\sigma^2_{xy}-\tilde\sigma^2_{xx}\left(\rho -\rho^{-1}\right)^2}}\right]\,,
\end{equation}
where $\tilde\sigma_{xx}\equiv\frac{\sigma_{xx}}{\rho}$.  The corresponding expression for the Kerr rotation angle is
\begin{equation}
  \theta_{K}= \frac{1}{2} Arg\left[ \frac{a\tilde\sigma_{xx}\left(\rho +\rho^{-1}\right)+i a \sqrt{4\sigma^2_{xy}-\tilde\sigma^2_{xx}\left(\rho -\rho^{-1}\right)^2}}{a\tilde\sigma_{xx}\left(\rho +\rho^{-1}\right)-i a \sqrt{4\sigma^2_{xy}-\tilde\sigma^2_{xx}\left(\rho -\rho^{-1}\right)^2}}\right]
\end{equation}
where we have omitted a negligible second order term $2 a^2 \left(\tilde\sigma_{xx} \tilde\sigma_{yy}+ \tilde\sigma^2_{xy}\right)$.

In Fig.~\ref{fig:image3}(b) and (c), we compare the Faraday and Kerr rotation angles calculated for the isotropic case ($\rho=1$, orange line) and the strongly anisotropic case ($\rho=0.1$, blue line).  The two plots are qualitatively similar, and their low-frequency limits do not depend on the anisotropy parameter $\rho$. However, the anisotropy results in a sharper enhancement of the Faraday rotation angle at the absorption edge, whereas the Kerr rotation angle is reduced. 

A more complete discussion of the Faraday and Kerr rotation angles in an anisotropic system is provided in Appendix A.

\section{Disorder broadening}

Our calculations thus far have been done with the assumption that momentum is strictly conserved, i.e., there is no impurity scattering. 
A simple, although nonrigorous way, to take into account the effect of a finite momentum relaxation time $\tau$ is to replace  the infinitesimal $\eta$ in  Eq.~(\ref{Current-CurrentResponse}) by the finite quantity $\delta=1/\tau$.  This is expected to be qualitatively correct as long  as $\delta$ remains much smaller than the gap $\Delta$. With this modification, the conductivity is calculated straightforwardly and works out to be
\begin{align}
         \sigma_{xy}(\omega) &= - \frac{ e^2}{4\pi\hbar}\frac{\Delta}{\omega}\log \frac{\omega'-\Delta}{\omega'+\Delta}\ \ ;\\
          \sigma_{xx}(\omega) &=\frac{i\rho e^2}{8\pi\hbar\omega} \left[\Delta+ \frac{(\omega'^2+\Delta^2)}{2\omega'}\ \log\frac{\omega'-\Delta}{\omega'+\Delta} \right]\,,
\end{align}
where $\omega' \equiv \omega + i \delta$ and $\log$ is understood to denote the complex logarithm.
From the expression of the conductivities above, we can see that the logarithmic divergence is not present anymore. Instead, we have a function that has a finite peak at the absorption edge.  The Faraday rotation angle is still given by Eq.~(\ref{theta_chiral2}), and it is plotted in Fig.~\ref{fig:image3}(d).  We see that with  increasing disorder, the curve is broadened, and the peak value is decreased. A significant enhancement  of $\theta_F$ can still be observed if the system is sufficiently pure.

\section{Conclusion}
 We have performed a detailed theoretical analysis of the Faraday and Kerr rotation for a generic model of gapped nonsymmorphic 2D semimetals. Both the gap and the required breaking of time-reversal symmetry are caused by a Zeeman term coupling with the out-of-plane component of the spin.  The Zeeman coupling could be generated by a real magnetic field, but such a field would also couple to the orbital motion of the electrons in the plane. 
 This is not the case for our model as it depends on the exchange interaction with a magnetic dopant or proximal magnetization -- the so-called proximity effect.  This gives rise to an "effective Zeeman coupling," which, being of Coulombic origin,  does not have the drawback of coupling to the orbital motion of the electrons in the same way that a real magnetic field does \cite{PhysRevLett.105.057401, PhysRevB.95.115405}.  

First, we matched Maxwell's equation for either side of the 2D layer to express the transmission and reflection matrices in terms of the conductivity matrix of the layer. We calculated the frequency-dependent  conductivity and finally computed the Faraday and Kerr rotation angles. We also observed the appearance of an elliptic component of the polarization, which inevitably accompanies the Faraday or the Kerr rotation when linearly polarized light is transmitted or reflected by the layer.

The Faraday rotation angle has a sharp peak at the absorption edge, which results in order of magnitude enhancement relative to its zero-frequency value.  We also find a giant Kerr rotation ($\approx \pi/2$) in the low-frequency region which abruptly changes signs just above the absorption edge.  The anisotropy of the system, parameterized by $\rho<1$, does not change these qualitative features but leads to an even sharper  enhancement of the Faraday rotation angle at the absorption edge. We also calculated the effect of a finite momentum relaxation time on the Faraday rotation angle due to disorder.  Not surprisingly, the peak becomes broader and weaker with increasing disorder. 

We propose that the dependence of the polarization rotation and the ellipticity on the incident angle, as shown, and the variation of their magnitudes with changing frequency, can be used  to study the optical properties of nonsymmorphic 2D Dirac semimetals experimentally in the presence of a gap-inducing effective Zeeman field.
The giant peak value in Faraday and Kerr rotation angles enabled by the band anisotropy and the field suggests that these materials can be useful platforms for optoelectronic and magnetoelectronic device applications.

\bibliography{mybib}
\newpage
\appendix
\section{Calculation of the polarization state for transmitted and reflected waves}
\begin{figure}
    \centering
    \includegraphics[width=0.4\linewidth]{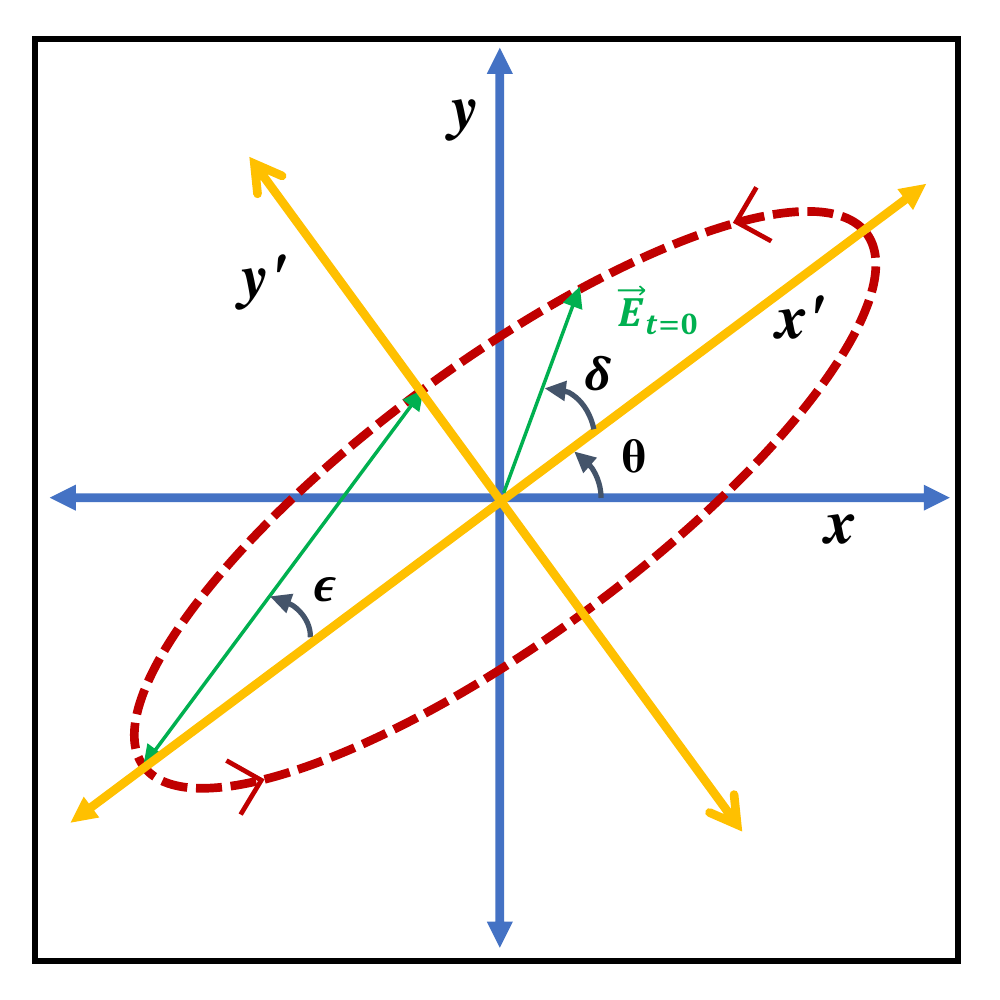}
\caption{\small Schematic diagram for left-handed elliptical polarization.  The major and minor axes are aligned along the $x'$ and $y'$ axes, respectively. The four parameters that define the ellipse of polarisation are (1) The angle $\theta$ between the major axis and a fixed axis of reference $x$. If the azimuthal angle $\theta$ is negative, it would be a right-handed polarization.(2) The ellipticity $e= b/a= \tan\epsilon$. The ellipticity has a range $-1 \leq e \leq 1$. If $e=-b/a=-\tan\epsilon$ instead, it is right-handed as well.(3) The total amplitude $A=\sqrt{a^2 +b^2}$ and (4) The absolute phase $\delta$ is defined as the angle between the electric field at $t=0$ and the major axis.}
    \label{fig:6}
\end{figure}

An elliptically polarized state with the major semiaxis oriented along the $x$ axis, with unit amplitude ($A=1$) and zero absolute phase ($\delta=0$) can be represented in the Cartesian basis ($x'-y'$ basis) (Fig.~\ref{fig:6}) \cite{AzzamR.M.A1977Eapl3} as follows:
\begin{equation}\label{refstate1}
    \Ev=\begin{pmatrix}
E_{x'}\\
E_{y'}\\
\end{pmatrix}=\begin{pmatrix}
\cos\epsilon\\
i \sin\epsilon\\
\end{pmatrix}\,,~~~-\pi/4<\epsilon<\pi/4\,,
\end{equation}
where $\epsilon$ is the ellipticity (for $\epsilon=0$ the polarization is linear along the $x$ axis, for $\epsilon=-\pi/4$ it is right-circular, for $\epsilon=\pi/4$ it is left-circular).

The general state of elliptic polarization has a major semi-axis ($x'$) that forms an angle $\theta$ with the $x$ axis. Thus the polarization state in the $x-y$ basis \footnote{Keeping the absolute phase of the electric field vector as zero ($\delta=0$)and of unit magnitude ($A=1$). One can multiply with $A e^{i\delta}$ to comprehend a more general scenario.} is obtained by applying a rotation by an angle $-\theta$ to the state~(\ref{refstate1}). This gives
\begin{equation}\label{ellipticalpol}
\begin{pmatrix}
E_{x}\\
E_{y}\\
\end{pmatrix}= R(-\theta)\begin{pmatrix}
E_{x'}\\
E_{y'}\\
\end{pmatrix}=\begin{pmatrix}
\cos\theta \cos \epsilon-i\sin\theta\sin\epsilon\\
\sin\theta \cos\epsilon+i \cos\theta\sin\epsilon\\
\end{pmatrix}\,,~~-\pi/4<\epsilon<\pi/4\,,~-\pi/2<\theta<\pi/2\,.
\end{equation}
Transforming to the chiral basis as we did before in Eq.~(\ref{Chiral_trans}) 
this becomes
\be\label{refstate21}
\begin{pmatrix}
E_{+}\\
E_{-}\\
\end{pmatrix}=
\frac{1}{\sqrt{2}}
\begin{pmatrix}
e^{i\theta}(\cos \epsilon-\sin\epsilon)\\
e^{-i\theta} (\cos\epsilon+ \sin\epsilon)\\
\end{pmatrix}=\begin{pmatrix}
z\\
z^{-1}\\
\end{pmatrix}\, 
\ee
where the complex number $z$ is given by
\ber
z&=& e^{i\theta}\sqrt{\frac{\cos \epsilon-\sin\epsilon}{\cos\epsilon+\sin\epsilon}}\nn\\
&=&e^{i\theta}\sqrt{\frac{1-\tan\epsilon}{1+\tan\epsilon}}\,.
\eer

Notice that ${\rm Arg}(z)=\theta$ varies between $-\pi/2$ and $\pi/2$, while $|z|=\sqrt{\frac{1-\tan\epsilon}{1+\tan\epsilon}}$ ranges from $0$ to $\infty$ as the ellipticity varies from $\pi/4$ to $-\pi/4$.
Thus it is easy to extract the parameters $\theta$ and $\epsilon$ from any given state vector expressed in the chiral representation.
Let us define a second measure of ellipticity $\gamma$ in the following manner 
\be\label{gammaellip}
e^{\gamma}=\sqrt{\frac{1-\tan\epsilon}{1+\tan\epsilon}}\,,~~~~\gamma=\frac{1}{2}\ln \frac{1-\tan\epsilon}{1+\tan\epsilon}
\ee
such that $\gamma=0$ for linear polarization and $\gamma=\infty$ for circular polarization. Then we can write
\be\label{Explicit-z}
z=e^{i(\theta-i\gamma)}\,.
\ee
In other words, $\gamma$ can be interpreted as the imaginary part of the angle that defines the orientation of the major semiaxis with respect to the $x$-axis.
Let us now consider the action of the transmission or reflection matrix on the incoming polarization state. Using the definition of $z$ from Eq.~(\ref{Explicit-z}) and acting with $\tilde \Tv$ (Eq.~\ref{Chiral_trans}) on the state~(\ref{refstate21}) we get the output 
\be
\tilde \Tv \cdot \begin{pmatrix}
e^{i\theta}e^{\gamma}\\
e^{-i\theta}e^{-\gamma}\\
\end{pmatrix} =\begin{pmatrix}
(\bar T+iT_{xy}) e^{i\theta}e^{\gamma}+\Delta T e^{-i\theta}e^{-\gamma}\\
(\bar T-iT_{xy}) e^{-i\theta}e^{-\gamma}+\Delta T e^{i\theta}e^{\gamma}\\
\end{pmatrix}
\ee 
This can be cast in the form $\left(\begin{array}{c} z'\\(z')^{-1}  \end{array}\right)$, where, following Eq.~(\ref{Explicit-z}), we have

\be
z' = e^{i\theta'}e^{\gamma'}=\sqrt{\frac{(\bar T+iT_{xy}) e^{i\theta}e^{\gamma}+\Delta T e^{-i\theta}e^{-\gamma}}{(\bar T-iT_{xy}) e^{-i\theta}e^{-\gamma}+\Delta T e^{i\theta}e^{\gamma}}}\,.
\ee
Finally, taking the complex logarithm and doing some simple transformations, we get
\be\label{complexfaraday}
\gamma'+i\theta'=\gamma+i\theta+\frac{1}{2} {\rm Ln}\frac{\bar T+iT_{xy} +\Delta T e^{-2i\theta}e^{-2\gamma}}{\bar T-iT_{xy} +\Delta T e^{2i\theta}e^{2\gamma}}
\ee 
Using Eq.~(\ref{complexfaraday}), we can find the complex angle, which has a real part, the Faraday angle, and an imaginary part, which is related to the ellipticity of the transmitted wave.  Separating the real and the imaginary parts of the logarithm we find

\begin{align}
\label{newellipticity}
\gamma'&=\gamma+\frac{1}{2}\ln \left\vert  \frac{\bar T+iT_{xy} +\Delta T e^{-2i\theta}e^{-2\gamma}}{\bar T-iT_{xy} +\Delta T e^{2i\theta}e^{2\gamma}}\right\vert\\
\label{newfaraday}
\theta'&=\theta+\frac{1}{2}{\rm Arg} \left(  \frac{\bar T+iT_{xy} +\Delta T e^{-2i\theta}e^{-2\gamma}}{\bar T-iT_{xy} +\Delta T e^{2i\theta}e^{2\gamma}}\right)\,.
\end{align}
The difference $\theta'-\theta$ is the Faraday rotation angle $\theta_{F}$, and we can calculate the ellipticity (=$\tan\epsilon$) using the Eq.~(\ref{gammaellip}) and from the difference $\gamma'-\gamma$. A similar treatment can be applied to the reflection matrix operating on the incoming polarization to get the Kerr rotation of reflected light.

In the isotropic case, $T_{xx}=T_{yy}=\bar T$ and $\Delta T=0$, the formula for the Faraday rotation angle reduces to 
\begin{equation}\label{theta_chiral}
   \theta'-\theta=\theta_F =\frac{1}{2} Arg \left[\frac{T_{xx}+ iT_{xy}}{T_{xx}- iT_{xy}}\right]=\frac{1}{2} Arg \left[\frac{T_{+}}{T_{-}}\right]\,,
\end{equation}
which agrees with Eq.~(\ref{theta_chiral2}) of the main text and is independent of the direction of polarization of the incident light. In the presence of anisotropy, however, the rotation angle does depend on the direction of polarization of the incident light (linear), and, in addition, the transmitted light is elliptically polarized. 

As a demonstration of the use of  Eqs.~(\ref{newellipticity},\ref{newfaraday}) and Eq.~(\ref{TransmissionMatrix2}), we  calculate the state of polarization of the outgoing wave for an incident wave that is  linearly polarized  along the $x$-direction (i.e., $\theta=0; \gamma=0$). We find (see also Eq.~\ref{ellipticalpol})) 
\begin{equation}\label{elliptic_out}
\begin{pmatrix}
E'_{x}\\
E'_{y}\\
\end{pmatrix}= \begin{pmatrix}
\cos\theta' \cos \epsilon'-i\sin\theta'\sin\epsilon'\\
\sin\theta' \cos\epsilon'+i \cos\theta'\sin\epsilon'\\
\end{pmatrix}=
\begin{pmatrix}
0.999748 - 0.000184 i\\
-0.020593 - 0.008917 i\\
\end{pmatrix}
\end{equation}
where  we have chosen $\rho=0.1$, the bandgap $\Delta=1$, and the frequency $\omega  = 1.1\Delta$. From Eq.~(\ref{elliptic_out}), we see  that an initially linearly polarized wave after passing through the medium is transformed into an elliptically polarized wave, with ellipticity $e=\tan\epsilon'\approx 0.01$.
\begin{figure}[t]
    \centering
 \includegraphics[width=0.50\linewidth]{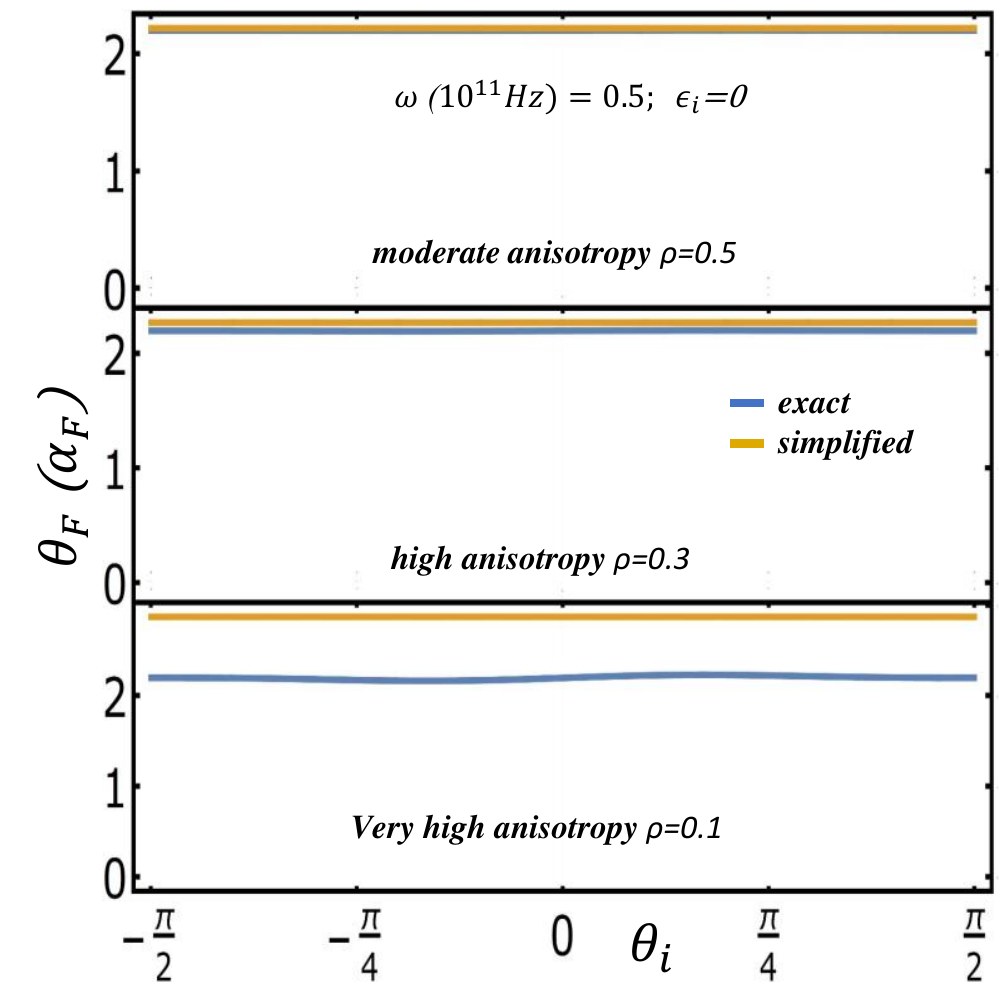}
 \caption{\small Faraday rotation as a function of the initial angle of linear polarization. The plots show the comparison between the exact and the simplified approach from our main text. Here the incident polarization is set to linear, making the ellipticity of the incident light zero ($\epsilon_{i}=0$). For a low anisotropy factor, the Faraday angle has a minimal difference between the two approaches. Still, with increasing anisotropy, the Faraday rotation has a mismatch between peak values.} 
    \label{fig:7}
\end{figure}

Now that we have established the exact formulation, let us benchmark the ``simplified" definition of the Faraday rotation angle we used in our main text.   For initial linear polarization ($\epsilon_i =0$), we find that the dependence of the Faraday rotation angle on $\theta_i$ is indeed negligible for all values of the anisotropy parameter (Fig.\ref{fig:7}).  In addition, the difference between the exact rotation angle and the approximate one of the main text is relatively small in a broad range of values of $\rho$, $0.1<\rho<1$. This remains true essentially for all frequencies.

\begin{figure}[t]
   \centering
    \includegraphics[width=1.0\linewidth]{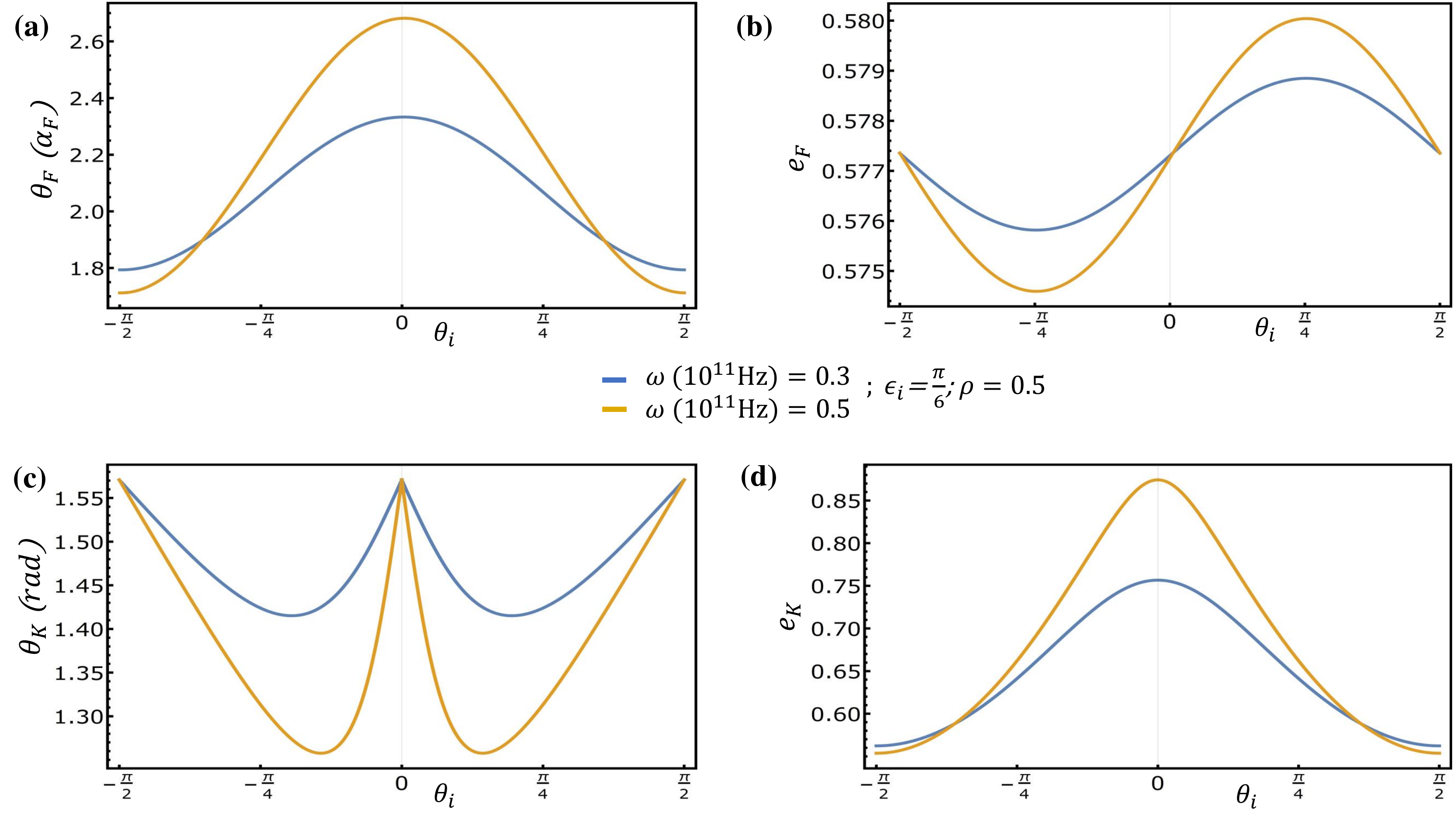}
\caption{\small Faraday and Kerr rotation as a function of the incident polarization angle and the optical frequency of the incident beam.
Panel (a) and (c) show the exact Kerr and Faraday rotation angles as 
functions of $\theta_{i}$ for $\omega=0.3$ and $\omega=0.5$, respectively. Panel (b) and (d) show the ellipticity $e  (= \tan\epsilon)$ of the transmitted and reflected wave as the function of $\theta_{i}$ for $\omega=0.3$ and $\omega=0.5$, respectively.} 
    \label{fig:8}
\end{figure}

Lastly, let us consider an elliptically polarized initial state and see what our exact formulas say. Fig.~(\ref{fig:8}) shows that the Faraday rotation angle -- and, more generally, the entire polarization state of the transmitted wave -- significantly depends on the initial angle $\theta_i$. Therefore, in this case (elliptically polarized incident light and strong anisotropy), the simplified approach we have used in the main text would not be accurate, and the exact formulas must be used instead. 

\end{document}